\documentstyle[prl,aps,multicol,epsf]{revtex}

\catcode`\@=11
\def\simle{\mathrel{\mathpalette\@versim<}}   
\def\simge{\mathrel{\mathpalette\@versim>}}   
\def\@versim#1#2{\lower2.5pt\vbox{\baselineskip0pt \lineskip-.5pt
   \ialign{$\m@th#1\hfil##\hfil$\crcr#2\crcr\sim\crcr}}}
\newcommand{\mib}[1]{\mbox{\boldmath $#1$}}
\newcommand{\mibs}[1]{\mbox{\boldmath {\scriptsize $#1$}}}

\newcommand{\citePR}[3]{Phys.\ Rev.\ {\bf #1}, #3 (#2)} 
\newcommand{\citePRB}[3]{Phys.\ Rev.\ B\ {\bf #1}, #3 (#2)} 
\newcommand{\citePRL}[3]{Phys.\ Rev.\ Lett.\ {\bf #1}, #3 (#2)}

\begin{document}

\draft

\title{
Two-dimensional charge order in layered 2-1-4 perovskite oxides
}
\author{
Shigeki Onoda$^1$, Yukitoshi Motome$^2$, and Naoto Nagaosa$^{1,3,4,5}$
}
\address{
$^{1}$Tokura Spin SuperStructure Project, ERATO,
Japan Science and Technology Corporation,\\
c/o Department of Applied Physics, University of Tokyo,
7-3-1, Hongo, Bunkyo-ku, Tokyo 113-8656, Japan
}
\address{
$^{2}$RIKEN (The Institute of Physical and Chemical Research),
2-1, Hirosawa, Wako, Saitama 351-0198, Japan
}
\address{
$^{3}$Department of Applied Physics, University of Tokyo,
7-3-1, Hongo, Bunkyo-ku, Tokyo 113-8656, Japan
}
\address{
$^{4}$Correlated Electron Research Center, National Institute of Advanced Industrial Science and Technology,\\
Tsukuba Central 4, 1-1-1 Higashi, Tsukuba, Ibaraki 305-8562, Japan
}
\address{
$^{5}$CREST, Japan Science and Technology Agency
}
\date{Recieved}

\maketitle
\begin{abstract}
Monte Carlo simulations are performed on the three-dimensional (3D) Ising model with the 2-1-4 layered perovskite structure as a minimal model for checkerboard charge ordering phenomena in layered perovskite oxides. Due to the interlayer frustration, only 2D long-range order emerges with a finite correlation length along the $c$ axis. Critical exponents of the transition change continuously as a function of the interlayer coupling constant. The interlayer long-range Coulomb interaction decays exponentially and is negligible even between the second-neighbor layers. Instead, monoclinic distortion of a tetragonal unit cell lifts the macroscopic degeneracy to induce a 3D charge ordering. The dimensionality of the charge order in La$_{0.5}$Sr$_{1.5}$MnO$_4$ is discussed from this viewpoint. 
\end{abstract}


\begin{multicols}{2}


Charge ordering (CO) phenomenon often appears in transition metal oxides 
including manganites, nickelates and probably cuprates. Of particular 
interest is the CO in layered perovskite oxides with K$_{2}$NiF$_{4}$ lattice structure (so-called 2-1-4 structure) such as 
La$_{2-x}$Sr$_xM$O$_4$ ($M$=Ni,Mn,Cu) 
\cite{Sternlieb1996,Yoshizawa2000,Kajimoto2001,Kajimoto2003,Tranquada1995}. 
When the carrier concentration is a rational number, a commensurate CO can 
occur. Particularly, in the case of $x=1/2$ or $3/2$, the CO within the 
$ab$ plane emerges with the checkerboard pattern at the wave vector 
${\mib Q} = (\pi/a, \pi/a)$, where $a$ is a lattice constant within the plane. 

Neutron scattering measurements have revealed that below the transition temperature $T_{\rm CO} = 217$ K, La$_{0.5}$Sr$_{1.5}$MnO$_4$ shows the 
checkerboard-type CO as a 2D long-range order (LRO) 
with a finite correlation length along the $c$ axis \cite{Sternlieb1996}. 
This provides an evidence of a Bragg rod. In contrast, recent X-ray 
\cite{Larochelle2001} and neutron \cite{Greven} scattering experiments have 
indicated a Bragg peak at a 3D wave vector $\tilde{\bf Q} = (\pi/a, \pi/a, 0)$ 
at a much lower temperature than $T_{\rm CO}$, indicating a 3D CO \cite{Greven}. The dimensionality of this CO in the material remains controversial. 

\begin{figure}
\epsfxsize=7cm
\centerline{\epsfbox{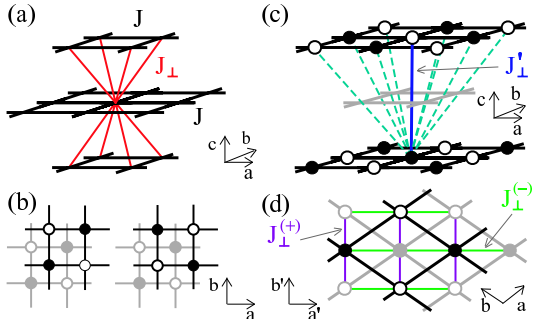}}
\caption{(Color)
(a) K$_{2}$NiF$_{4}$ tetragonal lattice structure and the interactions in the model (\ref{eq:H}). The interlayer coupling $J_\perp$ is partially drawn for the clarity of the figure.
(b) Frustrated stacking of two adjacent planes. The solid and gray lattices are for neighboring planes projected to the stacking direction. Open and filled circles denote up and down spins, respectively. 
(c) Second-neighbor interlayer couplings (blue solid line) and the longer-range interactions between the second-neighbor layers (light-blue dashed lines).
(d) Monoclinic lattice distortion to which the system shows an instability. 
Different interlayer couplings $J_\perp^{(\pm)}$ are introduced by this distortion.} 
\label{fig:model} 
\end{figure}

To resolve this controversy, the frustration of the interlayer Coulomb interaction shown in Fig.~\ref{fig:model} plays a central role: The two relative configurations of the neighboring 2D CO are completely degenerate when only the nearest neighbor interlayer coupling is taken into account. Frustrated interaction yields (i) complicated patterns of the ordering such as an incommensurate state, (ii) suppression of a LRO to realize the (quantum) liquid, (iii) glassy state, and so on \cite{Diep1994,Liebmann1986}. The degeneracy is usually lifted in a nontrivial way. Even a simple order emerges due to the so-called `order by disorder' mechanism \cite{Villain1980}, where the entropy of the fluctuation around each degenerate configuration differs and the system picks up the state with the largest entropy. 

Particularly for layered systems, when interlayer interaction suffers from a frustration, the dimensional reduction of a LRO occasionally occurs with a macroscopic degeneracy and only a weak universality relation is satisfied \cite{Popkov1997,Diep1994}: In special 3D models stacked with vertex models, a 2D LRO appears and critical exponents continuously vary with an interlayer coupling constant \cite{Popkov1997}, as in the eight-vertex model\cite{Baxter71,Wu71}. 
Therefore it is highly nontrivial and important to study what happens to the 
CO in the 2-1-4 structure in the presence of the interlayer frustration,
which we undertake in this paper. 
%


We model the CO in terms of the Ising model by assigning up and down spin configurations to Mn$^{3+}$ and Mn$^{4+}$ \cite{Zachar2003}. We consider the nearest-neighbor antiferromagnetic (AF) interaction in the 2-1-4 lattice structure shown in Fig.~\ref{fig:model} (a). The Hamiltonian is described as 
\begin{eqnarray}
{\cal H} &=& J \sum_{\ell} \sum_{\mibs{i}, \mibs{\eta}} \left(
  \sigma_{\mibs{i}}^{(2\ell)} \sigma_{\mibs{i}+\mibs{\eta}}^{(2\ell)}
+ \sigma_{\mibs{i}+\mibs{\delta}}^{(2\ell+1)}
  \sigma_{\mibs{i}+\mibs{\delta}+\mibs{\eta}}^{(2\ell+1)} \right)
\nonumber \\ 
&+& J_\perp \sum_{\ell} \sum_{\mibs{i}, \mibs{\eta}} \left(
  \sigma_{\mibs{i}}^{(2\ell)} \sigma_{\mibs{i}+\mibs{\delta}+\mibs{\eta}}^{(2\ell+1)}
+ \sigma_{\mibs{i}}^{(2\ell)}
  \sigma_{\mibs{i}+\mibs{\delta}-\mibs{\eta}}^{(2\ell+1)}
\right).
\label{eq:H}
\end{eqnarray}
Here $\sigma_{\mibs{i}}^{(\ell)} = \pm 1$ is the Ising variable defined at the site $\mib{i}$ in the $\ell$th layer. We have introduced 2D displacement vectors for nearest-neighbor sites $\mib{\eta} = (a,0)$ or $(0,a)$ within the plane and ${\mib \delta} \equiv (a/2, a/2)$ between the adjacent layers. Hereafter, we consider the case of $J > J_\perp$. In the limit of the independent 2D AF Ising models ($J_\perp = 0$), a phase transition takes place at $T_{\rm c}^{\rm 2D} = 2/(\sinh^{-1}1) \simeq 2.269$ \cite{Onsager1944}. Since $J_{\perp}$ is completely frustrated between the 2D AF ordered planes, there occurs a $2^{L_z}$-fold degeneracy corresponding to two-fold for each layer (Fig.~\ref{fig:model} (b)) where $L_z$ is the number of layers.

To study thermodynamic properties of the model (\ref{eq:H}) in detail, we perform Monte Carlo (MC) calculations. We employ the histogram algorithm \cite{LandauBinder} to obtain high-precision data, in addition to the metropolis algorithm with local flip as well as global or Wolff-type cluster flip \cite{LandauBinder}. We take $a=1$ as a length unit and $J=1$ as an energy unit hereafter.

\begin{figure}
\epsfxsize=8cm
\centerline{\epsfbox{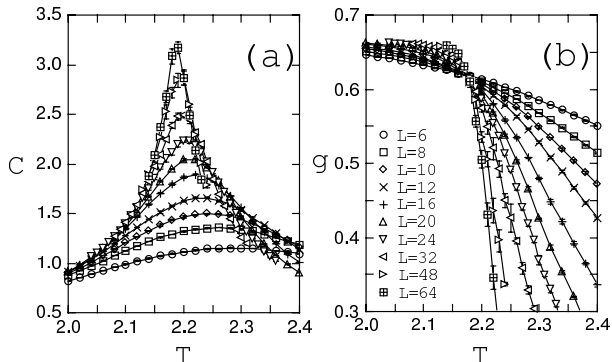}}
\caption{
Temperature dependences of (a) the specific heat and
(b) the Binder parameter in the case of $J_\perp = 0.5$.
The lines are guides to the eye.
}
\label{fig:transition}
\end{figure}

We expect that the AF order in each plane survives even with $J_\perp>0$. This is supported by MC results exemplified in Fig.~\ref{fig:transition} in the case of $J_\perp = 0.5$: A temperature dependence of the specific heat exhibits a singularity at $T \simeq 2.18$. At almost the same temperature, there occurs a systematic crossing of the Binder parameters $g$ \cite{LandauBinder} for the 2D antiferromagnetism at the different system sizes, namely, $g \equiv 1 - \langle M^4 \rangle / 3 \langle M^2 \rangle^2$ with the 2D AF order parameter for a certain layer $\ell$, $M = \sum_{\mibs{i}} \sigma_{\mibs{i}}^{(\ell)} {\rm e}^{{\rm i} \mibs{Q} \cdot \mibs{i}}$. The bracket denotes the thermal average for the canonical ensemble. These consistently indicate that at least the 2D AF LRO within each plane emerges at a critical temperature $T_{\rm c} \simeq 2.18$ which is slightly reduced from $T_{\rm c}^{\rm 2D}$. 

Usually, a 3D LRO due to the order-by-disorder mechanism \cite{Villain1980} 
is likely to occur because the entropy force drives 
a 3D ordering in the presence of a finite interlayer coupling. 
Despite this expectation, we found that there occurs no 3D 
ordering at all. Figure 3 shows the 3D spin structure factor 
$S({\bf k})$ at temperatures below $T_{\rm c}$ estimated from 
Fig.~\ref{fig:transition}. Here $S({\bf k})$ is defined as 
$ 
S({\bf k}) = \sum_{\ell\ell'} \sum_{\mibs{i},\mibs{j}} \langle \sigma_{\mibs{i}}^{(\ell)} \sigma_{\mibs{j}}^{(\ell')} \rangle {\rm e}^{{\rm i} 
{\bf k} \cdot ({\bf r}_{i\ell} - {\bf r}_{j\ell'})} / L^3
$ 
where ${\bf k}$ and ${\bf r}_{i\ell} = (\mib{i},\ell)$ are 3D vectors. Since the weight is concentrated on $S(\pi,\pi,k_z)$, we show the $k_z$ dependence of $S(\pi,\pi,k_z)/L^2$. There is no significant structure corresponding to the development of 3D Bragg peak. The inset shows a system-size dependence of $S(\pi,\pi,k_z)/L^3$, which clearly indicates that $S(\pi,\pi,k_z) \propto L^2$, i.e., the 2D LRO exists at every $k_z$ while there is no 3D ordering in the thermodynamic limit at all. Thus, there is a phase transition where the 2D AF correlation length diverges although the $c$-axis correlation length remains finite. 

This dimensional reduction is also confirmed by the low-temperature expansion of the free energy density with respect to $J_\perp$ \cite{AlexanderPincus}, 
$f=f_0+f_2+f_4+\cdots$ with 
$f_n\sim O((J_\perp\chi_{2D}(\mib{Q}))^n)$. 
$f_2$ and $f_4$ are calculated as
$f_2=\frac{T}{2L^2}\sum_{\mibs{k}}(4J_\perp\cos\frac{k_x}{2}
\cos\frac{k_y}{2})^2\chi_{2D}(\mib{k})^2$ 
and
$f_4 = \frac{T}{4!L^6} \sum_{\mibs{k}_1,\cdots,\mibs{k}_4} 
(\prod_{i=1,4}4J_\perp\cos\frac{k_{ix}}{2}
\cos\frac{k_{iy}}{2}) \tilde{\chi}_{2D} (\mib{k}_1,\mib{k}_2,\mib{k}_3,\mib{k}_4)^2$, respectively.
Here, $\chi_{2D}$ is the susceptibility for the 2D AF Ising model and 
$\tilde{\chi}_{2D}$ is a four-point susceptibility defined by 
$\tilde{\chi}_{2D} =
T^{-2}N^{-4}\sum_{\mibs{j}_1,\cdots,\mibs{j}_4}
\left(\prod_{i=1}^4e^{-{\rm i}\mibs{k}_i \cdot \mibs{j}_i}\right)
\times\Big[\langle\prod_{i=1}^4\sigma_{\mibs{j}_i}\rangle
-\sum_h{}'\langle\sigma_{\mibs{j}_{h(1)}}
\sigma_{\mibs{j}_{h(2)}}\rangle\langle
\sigma_{\mibs{j}_{h(3)}}
\sigma_{\mibs{j}_{h(4)}}\rangle\Big]$.
$\sum'$ represents the summation over 
three different combinations of $(1,2,3,4)$. 
The geometrical factor $\cos (k_x/2) \cos (k_y/2)$ 
appears at all orders in the expansion, which is 
a characteristic feature of the complete frustration of the present model. 
Furthermore, since $\chi_{2D}$ and $\tilde{\chi}_{2D}$ 
behave as $\exp(-\Delta/T)$ where $\Delta$ is the gap of the order of $J$, 
the expansion respect to $J_{\perp}$ is justified.
Therefore, there is no possibility of the 3D ordering at any wave vector.

Next, to understand the universality of this phase transition, we examine the critical exponents $\nu$ and $\eta$ in the case of $J_{\perp} = 0.5$. For this purpose, we perform MC simulations based on the histogram algorithm \cite{LandauBinder}, to obtain high-precision data on the following quantities in the vicinity of $T_{\rm c}$; 
the 2D spin structure factor 
$S_{\rm 2D}(\mib{Q}) = \langle M^2 \rangle / L^2$, 
the 2D spin susceptibility 
$\chi_{\rm 2D}(\mib{Q}) = \langle (M - \langle M \rangle )^2 \rangle$, 
$(d/d\beta)\log g$, 
$(d/d\beta)\log \langle M\rangle$ and 
$(d/d\beta)\log \langle M^2\rangle$.
We apply the finite-size scaling analysis to all the above quantities with
$S_{\rm 2D}(\mib{Q}) L^{\eta-2} = \Phi_1(\varepsilon L^{1/\nu})$, 
$\chi_{\rm 2D}(\mib{Q}) L^{\eta-2} = \Phi_2(\varepsilon L^{1/\nu})$,
$[(d/d\beta)\log g] L^{3-1/\nu} = \Phi_3(\varepsilon L^{1/\nu})$,
$[(d/d\beta)\log \langle M \rangle = \Phi_4(\varepsilon L^{1/\nu})$
and $[(d/d\beta)\log \langle M^2 \rangle = \Phi_5(\varepsilon L^{1/\nu})$.
Here, $\Phi$'s are universal functions and 
$\varepsilon = (T-T_{\rm c})/T_{\rm c}$ is a reduced temperature. 
We determine $T_{\rm c}$, $\nu$, $\eta$ and $\Phi_i$'s from the least-square 
fittings of all the data at different $L$ and $T$ to $\Phi_i$ with 
$i=1,\cdots,5$. Figure~\ref{fig:scaling} shows that this finite-size-scaling 
analysis works quite well without any logarithmic correction. The analysis gives $T_{\rm c}=2.177953(7)$, 
$\nu=0.8687(1)$ and $\eta=0.24575(6)$. Statistical errors in the last digit are shown in parentheses. This high precision is comparable to the best one obtained for
the conventional 3D Ising model \cite{LandauBinder}. The set of these values 
for critical exponents is not consistent with either the 2D Ising 
($\eta = 1/4$ and $\nu = 1$) \cite{Kaufmann1944} or the 3D Ising 
universality class ($\eta = 0.032$ and $\nu = 0.63$) \cite{LeGuillou1977}. 
This suggests that the present phase transition does not belong to either 
the 2D or the 3D Ising universality class.

For various values of $J_\perp$, we perform a finite-size-scaling analysis of $S_{\rm 2D}({\mib Q})$ based on the MC results of the metropolis algorithm \cite{Motome2001}. Figure~\ref{fig:Tc and exponents} summarizes the estimates of $T_{\rm c}$ and the critical exponents as a function of $J_\perp$: 
Increasing $J_\perp$ from $0$, $T_{\rm c}$ decreases continuously from $T_{\rm c}^{\rm 2D}$ and the exponent $\nu$ decreases continuously from the 2D Ising value $\nu = 1$, while the exponent $\eta$ remains similar to the 2D Ising value $\eta\sim 1/4$. Such continuous change of critical exponents appears in the eight-vertex model \cite{Baxter71,Wu71} and Ashkin-Teller model \cite{AshkinTeller43} and the so-called sliding phase of a 3D XY model \cite{OHernLubenskyToner99}. We also found that in the bilayer system, the continuously varying $\nu$ but with almost fixed $\eta$ is obtained (Fig.~\ref{fig:Tc and exponents}). The universality class of this 2D phase transition is not determined by a fixed point but by a fixed line, which is also found in some 2D frustrated systems \cite{Popkov1997}. Namely, the interlayer coupling between the adjacent layers modifies the universality class of the model.


To understand the dimensionality of the checkerboard CO experimentally observed in La$_{0.5}$Sr$_{1.5}$MnO$_4$, we discuss here the stability of the 2D LRO against possible 3D LRO's due to (I) the long-range Coulomb interaction between layers and (II) lattice distortion. 

(I) First, we add to the model (\ref{eq:H}) the following interlayer 
coupling term between the second-neighbors (Fig.~\ref{fig:model} (c))
to mimic the longer-range Coulomb interaction;
$
{\cal H}' = J'_\perp \sum_{\ell} {\cal O}_{\rm I}
$
with 
$
{\cal O}_{\rm I}^{(\ell)}\equiv\sum_{\mibs{i}} \sigma_{\mibs{i}}^{(\ell)} 
\sigma_{\mibs{i}}^{(\ell+2)}.
$
The susceptibility to $J'_\perp$ is given by 
$\chi_4 = \langle {{\cal O}_{\rm I}^{(\ell)}}^2 \rangle / L^2$. 
If $\chi_4$ is finite for $L \to \infty$, the 2D order is stable at 
least for small $J_\perp'$ as in low-dimensional ordered phases in liquid 
crystals\cite{chaikin}. On the contrary, our MC results for $\chi_4$ show 
a divergence as $L^2$ in the limit of $L \to \infty$. This means that the 
spins within each layer act as a big spin of the size of $L^2$. Then, the 
present 2D LRO is unstable against a $J_\perp'$-driven 3D LRO.  
However, this effect of $J_\perp'$ is almost cancelled out by the remaining part of $1/r$ longer-range Coulomb interaction than $J_\perp'$ between different layers: 
We define the average spin in the $\ell$th layer as 
$S^{(\ell)}=L^{-2}\sum_{\mibs{i}}\sigma^{(\ell)}_{\mibs{i}}e^{{\rm i}\mibs{Q}\cdot\mibs{i}}$ 
and an effective interaction $J_\perp^{\rm eff}(\ell-\ell')$ between 
$S^{(\ell)}$ and $S^{(\ell')}$. We note $J_\perp^{\rm eff}(2\ell+1)=0$. 
There remains the complete frustration between the odd-spaced layers. 
Using the Ewald summation, $J_\perp^{\rm eff}(2\ell)$ is found to decay 
as $e^{-4.5\ell c/a}$ for $c/a\simge 1$. Then even for the
second-neighbor layers, $J_\perp^{\rm eff}(2)/J$ is reduced to $10^{-8}$ 
with the lattice constants $a=3.86\AA$ and $c=12.44\AA$ \cite{Bouloux1981}. 
This extremely small energy scale of the order of $\mu$K should be
irrelevant in realistic experimental situations.

(II) In realistic materials, there exists the electron-lattice coupling. 
In the present model in Eq. (\ref{eq:H}), if the lattice is deformed, 
the couplings $J$ and/or $J_\perp$ are modified. There are many candidates 
for the lattice distortion to lift the macroscopic degeneracy accompanied 
with the 2D LRO. Up to the linear order in distortion, however, the system 
is unstable against only a monoclinic distortion of the tetragonal unit 
cell (Fig.~\ref{fig:model} (d)). This distortion produces two different 
interlayer couplings $J_\perp^{(\pm)}$ so that the frustration is partly 
removed. It is useful to introduce a structure factor corresponding to 
this distortion as $F \equiv \langle {\cal O}^2_{\rm II}\rangle/L^3$, where 
$
{\cal O}_{\rm II} \equiv \sum_{\mibs{i},\ell}\sum_{s=\pm1}
 \sigma_{\mibs{i}}^{(2\ell)}
(\sigma_{\mibs{i}+\mibs{\delta}^{(+)}}^{(2\ell+s)}
+\sigma_{\mibs{i}-\mibs{\delta}^{(+)}}^{(2\ell+s)}
-\sigma_{\mibs{i}+\mibs{\delta}^{(-)}}^{(2\ell+s)}
-\sigma_{\mibs{i}-\mibs{\delta}^{(-)}}^{(2\ell+s)})
$
with $\mib{\delta}^{(\pm)}=(1/2, \pm1/2)$ is the operator that directly couples to the distortion. In the case of $J_\perp=0.5$, from the finite-size scaling of MC results shown in Fig.~\ref{fig:F}, we found that $F$ diverges toward $T_{\rm c}$ in the thermodynamic limit. The 2D order is unstable against the monoclinic distortion to lead a 3D LRO. If we rotate the unit cell by $45^\circ$ around the $c$ axis, this distortion represents an orthorhombic one. This is achieved by applying a uniaxial pressure along the $a'$ or $b'$ axis in the rotated frame.

In conclusion, we have studied the phase transition and the dimensionality of the LRO in the AF Ising model on the 2-1-4 lattice: If the crystal structure has an ideal tetragonal symmetry, the 2D LRO appears with the nontrivial critical phenomena at $T_{\rm c}$. On the other hand, the system is unstable towards the
monoclinic distortion, which replaces the 2D LRO with a 3D LRO. 
This conclusion applies to the checkerboard CO phenomenon in 
La$_{0.5}$Sr$_{1.5}$MnO$_4$. In the former case without the distortion, one should observe a 
stronger singularity in the specific heat with the exponent 
$\alpha = 2(1-\nu) > 0$ than the logarithmic divergence. 
In the distorted case, a change of the lattice constants 
$b_{\rm o}\ne2a_{\rm o}$ in the orthorhombic unit cell should be 
observed as an additional feature to a CO-induced doubling of a unit cell 
and possibly a proposed pattern of distortion related to orbital 
order \cite{Wilkins2003}. 

The authors acknowledge J. Zaanen, S. Sachdev, Y. Tokura, S. Miyashita, 
H. Tsunetsugu and N. Kawakami for useful discussions. 
This work is supported by Grant-in-Aids from the Ministry of Education,  
Culture, Sports, Science, and Technology. 


\begin{figure}
\epsfxsize=7cm
\centerline{\epsfbox{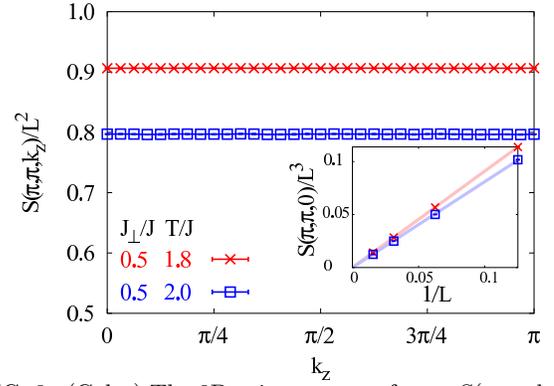}}
\caption{(Color)
The 3D spin structure factor $S(\pi,\pi,k_z)$ divided by the square dimension $L^2$ is plotted against $k_z$ below $T_{\rm c}$ in the case of $L=64$. The inset shows that $S(\pi,\pi,0)/L^3$ linearly goes to zero with $1/L$.} 
\label{fig:S(pi,pi,k_z)} 
\end{figure}

\begin{figure}
\epsfxsize=8.5cm
\centerline{\epsfbox{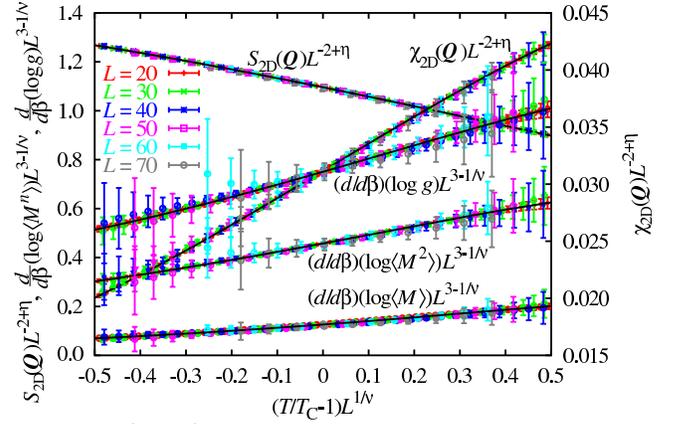}}
\caption{(Color)
The best-fit result of the finite-size scaling for correlation functions 
obtained by our MC calculation based on the histogram algorithm in the case of $J_\perp = 0.5$.} 
\label{fig:scaling} 
\end{figure}

\begin{figure}
\epsfxsize=7cm
\centerline{\epsfbox{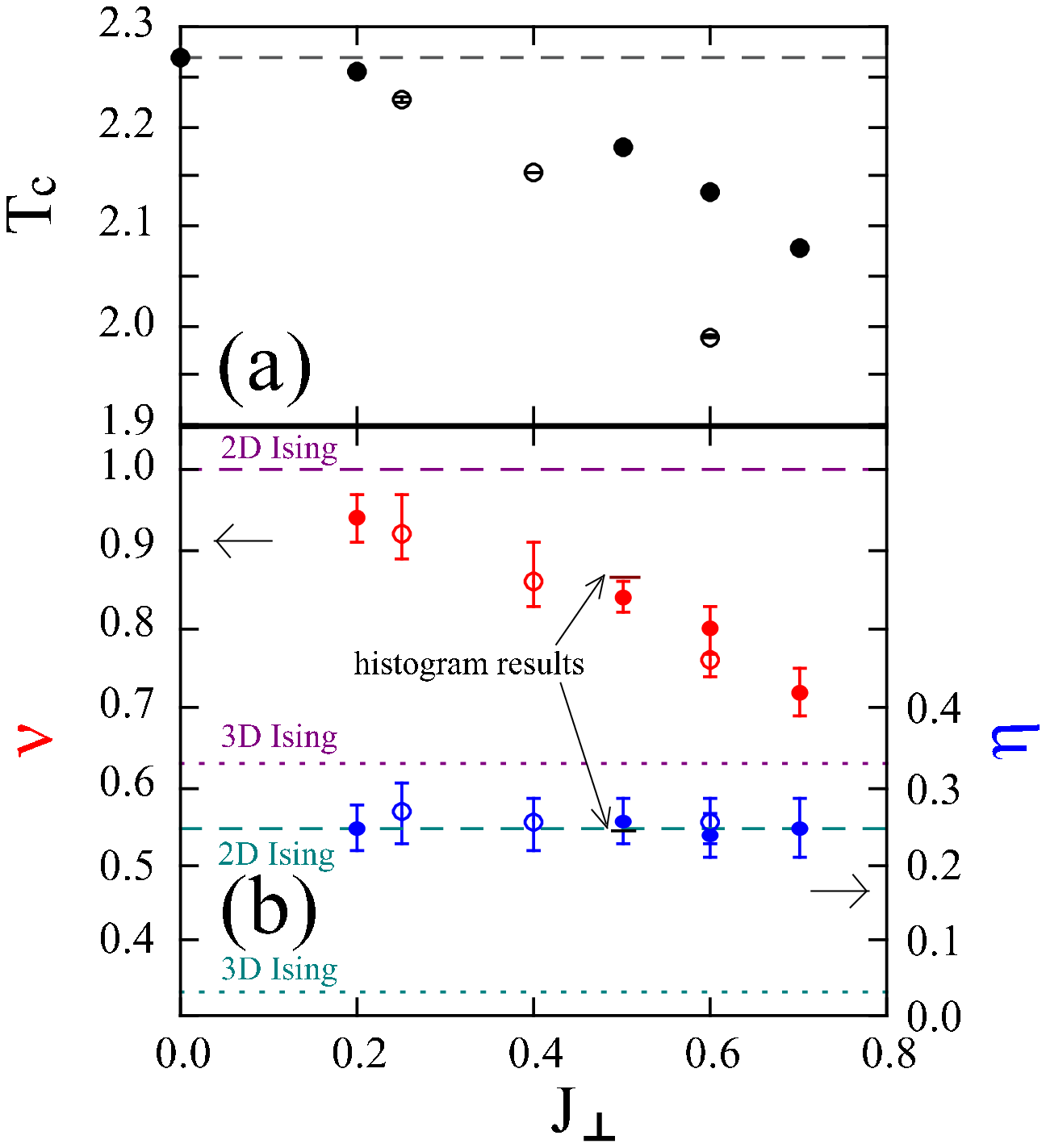}}
\caption{(Color)
$J_\perp$ dependences of (a) the critical temperature,
(b) the critical exponents $\nu$ and $\eta$.
Dashed (dotted) lines denote the values of the 2D (3D) Ising model.
Filled (open) symbols are for $L^3$ 
($L^2 \times 2$, namely bilayer) systems
in model (\ref{eq:H}).
Errorbars are omitted when they are smaller than the symbol sizes.
Note that the values of $J_\perp$ for 2-layer systems
are divided by factor of $2$ for convenience.
}
\label{fig:Tc and exponents}
\end{figure}

\begin{figure}
\epsfxsize=7.5cm
\centerline{\epsfbox{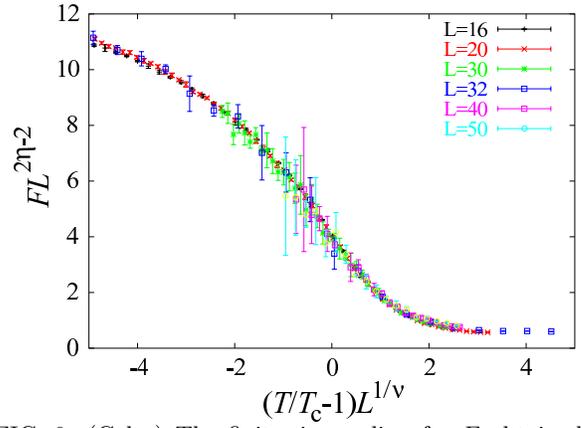}}
\caption{(Color) 
The finite-size scaling for $F$ obtained by our MC calculation in the case of $J_\perp = 0.5$. We have used the values of $T_{\rm c}$ and $\nu$ obtained from the analysis given in Fig.~\ref{fig:scaling}.} 
\label{fig:F}
\end{figure}

\end{multicols}

\begin{thebibliography}{99}

\bibitem{Sternlieb1996}
B. J. Sternlieb {\it et al.}, Phys. Rev. Lett. {\bf 76}, 2169 (1996).

\bibitem{Yoshizawa2000}
H. Yoshizawa {\it et al.}, Phys. Rev. B {\bf 61}, R854 (2000).

\bibitem{Kajimoto2001}
R. Kajimoto {\it et al.}, Phys. Rev. B {\bf 64}, 144432 (2001).

\bibitem{Kajimoto2003}
R. Kajimoto {\it et al.}, Phys. Rev. B {\bf 67}, 014511 (2003).

\bibitem{Tranquada1995}
J.M. Tranquada {\it et al}., Nature (London) {\bf 375}, 561 (1995).

\bibitem{Larochelle2001}
S. Larochelle {\it et al.}, Phys. Rev. Lett. {\bf 87}, 095502 (2001).

\bibitem{Greven}
M. Greven: private communications.

\bibitem{Diep1994}
For instance, see {\it Magnetic System with Competing Interaction}, ed. by 
H.T. Diep (World Scientific Publishing Co., 1994).

\bibitem{Liebmann1986}
R. Liebmann, {\it Statistical mechanics of periodic frustrated Ising systems} 
(Springer-Verlag, Berlin, Tokyo, 1986).

\bibitem{Villain1980}
J. Villain {\it et al.}, J. Phys. (Paris) {\bf 41}, 1263 (1980).

\bibitem{Popkov1997}
H.Y. Huang, V. Popkov, and F.Y. Wu, Phys. Rev. Lett. {\bf 78}, 409 (1997). 
Similar dimensional reduction was also found in 2D systems only at $T=0$ \cite{Diep1994}. However, in contrast with the 1D LRO, the 2D LRO is allowed even at finite temperatures where the 2D LRO may be replaced with a 3D order due to the order-by-disorder mechanism.

\bibitem{Baxter71}
R.J. Baxter, \citePRL{26}{1971}{832}.

\bibitem{Wu71}
F.W. Wu, \citePRB{4}{1971}{2312}.

\bibitem{Zachar2003}
O. Zachar and I. Zaliznyak, Phys. Rev. Lett. {\bf 91}, 036401 (2003).

\bibitem{Onsager1944}
L. Onsager, Phys. Rev. {\bf 65}, 117 (1944).

\bibitem{LandauBinder}
D.P. Landau and K. Binder, {\it A Guide to Monte Carlo Simulations in Statistical Physics} (Cambridge University Press, 2000).

\bibitem{AlexanderPincus}
S. Alexander and P. Pincus, J. Phys. A {\bf 13}, 263 (1980).

\bibitem{Kaufmann1944}
B. Kaufmann and L. Onsager, Phys. Rev. {\bf 76}, 1244 (1944);
J. Stephenson, J. Math. Phys. {\bf 5}, 1009 (1964).

\bibitem{LeGuillou1977}
J.C. LeGuillou and J. Zinn-Justin, Phys. Rev. Lett. {\bf 39}, 95 (1977).

\bibitem{Motome2001}
For details of fitting and error estimation, see
Y. Motome and N. Furukawa, J. Phys. Soc. Jpn. {\bf 70}, 1487 (2001).

\bibitem{AshkinTeller43}
J. Ashkin and E. Teller, \citePR{64}{1943}{178}.

\bibitem{OHernLubenskyToner99}
In the sliding phase of a 3D XY model [C.S. O'Hern, T.C. Lubensky and J. Toner, \citePRL{83}{1999}{2745}], only a 2D quasi LRO is allowed at $T>0$. Then, the scaling dimension of an interlayer coupling is not enough to establish a 3D LRO. The interlayer coupling modifies critical exponents as in our case. However, this quasi 2D LRO is quite different from our 2D LRO.

\bibitem{chaikin}
In liquid crystals, the density of molecules has 1D LRO in the smectic phase and 2D LRO in the discotic phase. It is disordered in the other remaining direction. Then, the Bragg plane and rod appears, respectively. [P.M. Chaikin and T.C. Lubensky, {\it Principles of condensed matter physics} (Cambridge University Press, 1995)].

\bibitem{Bouloux1981}
J.C. Bouloux {\it et al}., 
Mater. Res. Bull. {\bf 16}, 855 (1981).

\bibitem{Wilkins2003}
S.B. Wilkins {\it et al}., \citePRL{91}{2003}{167205}.

\end{thebibliography}
\end{document}